\documentclass{elsart}
\usepackage{amsmath}

\begin{document}
\begin{frontmatter}

\title{Nonintegrability of (2+1)-dimensional continuum isotropic Heisenberg spin
system: Painlev\'e analysis}

\author{C.~Senthil~Kumar$^{a}$},
\author{M.~Lakshmanan$^{a,}$\thanksref{mail2}},
\author{B.~Grammaticos$^{b}$} and
\author{A.~Ramani$^{c}$} 

{\footnotesize$^a$}{\it Centre for Nonlinear Dynamics, Dept. of Physics, Bharathidasan University,
 Tiruchirapalli - 620 024, India. }\\
{\footnotesize$^b$}{\it GMPIB, Université Paris VII, Tour 24-14, 5eétage, case 7021, 75251 Paris, France.} \\
{\footnotesize$^c$}{\it CPT, Ecole Polytechnique, CNRS, UMR 7644, 91128 Palaiseau, France.}
\thanks[mail2]{{\it Corresponding author:  E-mail: }\;\;lakshman@cnld.bdu.ac.in}
\date{}

\begin{abstract}

While many integrable spin systems are known to exist in (1+1) and (2+1)
dimensions, the integrability property of the physically important (2+1)
dimensional isotropic Heisenberg ferromagnetic spin system in the continuum
limit has not been investigated in the literature.  In this paper, we show
through a careful singularity structure analysis of the underlying nonlinear
evolution equation that the system admits logarithmic type singular manifolds
and so is of non-Painlev\'e type and is expected to be nonintegrable. 
\end{abstract}

\begin{keyword}
Painlev\'e property; Integrability
\PACS{02.30.Jr; 02.30.Ik; 75.10.Pq }

\end{keyword}

\end{frontmatter}

The nonlinear dynamics underlying magnetic spin systems
is a fascinating topic of study and it is of considerable interest especially from the
points of view of soliton theory and condensed matter physics.
The underlying evolution equations are highly nonlinear and they give rise to
many integrable cases both in (1+1) and (2+1) dimensions.

The standing example of an integrable spin system in (1+1)
dimensions is the isotropic Heisenberg ferromagnetic spin (IHFS)
chain \cite{ref1}-\cite{ref3} in its continuum limit.  The
underlying spin evolution equation is
\begin{equation}
{\bf S}_{t}={\bf S}\wedge{\bf S}_{xx}, \label{1}
\end{equation}
where ${\bf S}=(S_1,S_2,S_3),\quad {\bf S}^2=1$. It is equivalent
geometrically \cite{ref2} and through gauge transformation
\cite{ref3} to the ubiquitous soliton possessing nonlinear
Schr$\ddot{o}$dinger equation \cite{ref2}. Also the corresponding
spin evolution equation itself is associated with a Lax pair and
the inverse scattering transform analysis can be carried out for
the system directly  \cite{ref3_1}.

Besides the isotropic spin system, there exists a number of other
spin systems in (1+1) dimensions which possess Lax pairs, gauge
equivalent counterparts and complete integrability property. These
include the addition of anisotropy and magnetic field to the
isotropic case leading to the spin evolution equation \cite{ref4},
\begin{equation}
{\bf S}_{t}={\bf S}\wedge{\bf S}_{xx}-2A({\bf S}.{\bf n}){\bf n}+\mu{\bf B},
\end{equation}
where $ \vec{n}=(0,0,1),$ $ \vec{B}=(0,0,B)$, A is the strength of
anisotropy and B is the strength of the magnetic field along the
$z$-direction.

One more interesting integrable spin evolution equation is the
bianisotropic equation studied by Sklyanin \cite{ref5},
\begin{equation}
 {\bf S}_{t}={\bf S}\wedge {\bf S}_{xx}+{\bf S}\wedge{\bf J}{\bf S},
\end{equation}
where $ {\bf J}=diag(J_1,J_2,J_3)$ is the anisotropic matrix. The
above type of spin equations are also special cases of the
Landau-Lifshitz(L-L) equation deduced from phenomenological
arguments \cite{ref5_1}. Besides the aforementioned systems,
various higher order and inhomogeneous integrable extensions
 also exist. For example, the spin evolution equation
\begin{equation}
{\bf S}_{t}=(\nu_2+\mu_2 x){\bf S}\wedge
{\bf S}_{xx}+\mu_2 {\bf S}\wedge{\bf S}_{x}-(\nu_1+\mu_1 x){\bf S}_{x}
-\gamma [{\bf S}_{xx}+
{3 \over{2}}({\bf S}_{x})^2.
{\bf S}]_x,
\end{equation}
is integrable \cite{ref6}. Also an SO(3) invariant deformed Heisenberg spin
system has been shown to be integrable \cite{ref7}
\begin{equation}
{\bf S}_{t}={\bf S}\wedge{\bf S}_{xx}+\alpha{\bf S}_{x} ({\bf
S}_{x})^2,
\end{equation}
and it is equivalent to the integrable derivative NLS equation \cite{ref8}
\begin{equation}
iq_t+q_{xx}+ 2|q|^2q-2i\alpha(|q|^2q)_x = 0.
\end{equation}

Many other integrable generalizations have also been obtained by
Myrzakulov and coworkers \cite{ref9}-\cite{ref12}. All the above equations admit Lax
pairs and satisfy Painlev\'e property.

Naturally, the question arises as to what is the situation in
(2+1) dimensions.  The well known integrable generalization of Eq.
(\ref{1}) in (2+1) dimensions are the Ishimori equation \cite{ref13},
\begin{subequations}
\begin{eqnarray}
{\bf S}_{t}={\bf S}\wedge({\bf S}_{xx}+\sigma^2{\bf S}_{yy})+\phi_y {\bf S}_x+\phi_x {\bf S}_y,\\
\phi_{xx}-\sigma^2\phi_{yy}=-2\sigma^2 {\bf S}.{\bf S}_x\wedge{\bf
S}_y,
\end{eqnarray}
\end{subequations}
where ${\bf S}=(S_1,S_2,S_3),\quad {\bf S}^2=1$ and $\phi(x,y,t)$
is a scalar field and $\sigma^2=\pm 1$, and the Myrzakulov M-I 
equation \cite{ref11}
\begin{subequations}
\begin{eqnarray}
{\bf S}_{t}=\{{\bf S}\wedge({\bf S}_{y}+u{\bf S}\}_x,  \\
u_x=-{\bf S}.{\bf S}_x\wedge{\bf S}_y,
\end{eqnarray}
\end{subequations}
where $u(x,y,t)$ is a scalar field.

Again these equations possess Lax pairs and admit the Painlev\'e
property.  However, till today the integrability nature of the
physically interesting (2+1) dimensional direct generalization of (\ref{1}),
namely
\begin{equation}
{\bf S}_{t}={\bf S}\wedge({\bf S}_{xx}+{\bf S}_{yy}), \label{2}
\end{equation}
where ${\bf S}=(S_1,S_2,S_3),\quad {\bf S}^2=1$, has not been
studied, though the special case of Eq.(\ref{2}) with circular symmetry
\begin{equation}
{\bf S}_{t}={\bf S}\wedge({\bf S}_{rr}+{1 \over{r}}{\bf S}_{r}), \label{3}
\end{equation}
where $ r=\sqrt{x^2+y^2}$, is known to be integrable \cite{ref14}.

In this paper, we wish to investigate the singularity structure
property of the isotropic Heisenberg spin equation (\ref{2}) in
(2+1) dimensions and prove that it is of non-Painlev\'e type and
so is expected to be non-integrable, even though the special cases
(\ref{1}) and (\ref{3}) are of Painlev\'e type and so integrable.
The Painlev\'e analysis of the Heisenberg spin type equations is rather
tricky as was shown for the case of (1+1) dimensional system with
anisotropy and transverse magnetic field \cite{ref15}, where the "Taylor"
type expansion can lead to logarithmic singular manifolds leading
to non-integrability.

In order to investigate the Painlev\'e singularity structure
underlying Eq. (\ref{2}), we first rewrite it in terms of the
complex stereographic field variable $\omega(x,y,t)$ through the
transformation
\begin{equation}
S^+=S_1+iS_2=\frac{2\omega}{1+|\omega|^2},
S_3=\frac{1-|\omega|^2}{1+|\omega|^2}.
\end{equation}
In terms of this variable, the equation of motion for the (2+1) dimensional Heisenberg spin system
can be written as
\begin{equation}
(1+\omega\omega^*)[i\omega_{t}+\omega_{xx}+\omega_{yy}]-2\omega^*
(\omega_{x}^2+\omega_{y}^2)=0, \label{4}
\end{equation}
and its complex conjugate.  Representing $\omega \rightarrow F$ and $\omega^* \rightarrow G$,
Eq. (\ref{4}) and its complex conjugate equation can be written as
\begin{subequations}
\begin{eqnarray}
(1+FG)(iF_{t}+F_{xx}+F_{yy})-2G(F_{x}^2+F_{y}^2)=0, \\
(1+FG)(-iG_{t}+G_{xx}+G_{yy})-2F(G_{x}^2+G_{y}^2)=0.
\end{eqnarray} \label{5}
\end{subequations}
We carry out a Painlev\'e analysis of Eqs.(\ref{5}) by seeking a
generalized Laurent expansion for each dependent variable in the
form,
\begin{subequations}
\begin{eqnarray}
F=F_{0}\phi^p+\sum_{j}F_{j}\phi^{p+j},\;\;\; F_{0}\neq0,\\
G=G_{0}\phi^q+\sum_{j}G_{j}\phi^{q+j}, \;\;\; G_{0}\neq0,
\end{eqnarray}\label{6}
\end{subequations}
in the neighbourhood of the noncharacteristic singular manifold
$\phi(x,y,t)=0$, $\phi_t, \phi_x, \phi_y \neq 0$. The results are
as follows.\\ 
\newline
{\it 1.Leading Order Behaviour}

Looking at the dominant terms, we distinguish the following
possibilities corresponding to (i) $p\leq0$, $q\leq0$, (ii)
$p\leq0$, $q\geq0$, (iii) $p\geq0$, $q\leq0$.
\\ \\
{\it Case(i): $p\leq0$, $q\leq0$} :

Upon using the leading order solution $F=F_{0}\phi^p$,
$G=G_{0}\phi^q$, substituting it in Eq.(\ref{5}), and balancing the
most dominant terms, we obtain
\begin{subequations}
\begin{eqnarray}
F_{0}^2G_{0}[p(p-1)-2p^2](\phi_{x}^2+\phi_{y}^2)\phi^{2p+q-2}=0, \\
F_{0}G_{0}^2[q(q-1)-2q^2](\phi_{x}^2+\phi_{y}^2)\phi^{2q+p-2}=0.
\end{eqnarray}
\end{subequations}
From the above, we have the following three possibilities of leading order
behaviour:
\\

\begin{tabular}{p{2.2cm}p{3cm}p{8cm}}
Branch (i)  &  $p=-1,q=-1,$ & $F_{0}$, $G_{0}$: arbitrary\\
Branch (ii) &  $p=-1,q=0, $ & $F_{0}$, $G_{0}$: arbitrary\\
Branch (iii)&  $p=0,q=-1, $ & $F_{0}$, $G_{0}$: arbitrary
\end{tabular}
\\ \\
In addition, there is a possibility that $p=0$,
$q=0$, which requires a more detailed analysis, see below.
\\
\newline
{\it Case(ii): $p\leq0$, $q\geq0$:}
\begin{subequations}
\begin{eqnarray}
\big(F_{0}(p-1)\phi^{p-2}-F_{0}^2G_{0}(p+1)\phi^{2p+q-2}\big)(\phi_{x}^2+\phi_{y}^2)=0, \label{7a} \\
\big(G_{0}(q-1)\phi^{q-2}-F_{0}G_{0}^2(q+1)\phi^{p+2q-2}\big)(\phi_{x}^2+\phi_{y}^2)=0. \label{7b}
\end{eqnarray}
\end{subequations}

From Eqs. (\ref{7a},\ref{7b}) we obtain $p+q=0$ and
$F_{0}G_{0}=\frac{p-1}{p+1}$ from Eq.(\ref{7a}), and
$F_{0}G_{0}=\frac{q-1}{q+1}$ from Eq.(\ref{7b}), respectively.  We
also obtain the same result for the case $p\geq0,q\leq0$. This
suggests that $p=q=0$ is the only possibility here.  Looking at
this case more carefully, by using Eq. (\ref{6}) in (\ref{5}), we
obtain the following.

Branch (iv): $p=0$, $q=0$
\begin{subequations}
\begin{eqnarray}
(1+F_{0}G_{0})[i(F_{0t}+F_{1}\phi_{t})+F_{0xx}+2F_{1x}\phi_{x}+F_{1}\phi_{xx}+2F_{2}
\phi_{x}^2+F_{0yy} \nonumber \\
+2F_{1y}\phi_{y}+F_{1}\phi_{yy}+2F_{2}\phi_{y}^2]
-2G_{0}[F_{0x}^2+F_{0y}^2+F_{1}^2(\phi_{x}^2+\phi_{y}^2)\nonumber \\
+2F_{1}(F_{0x}\phi_{x}+F_{0y}\phi_{y})]=0,\label{7_1a}\\
(1+F_{0}G_{0})[-i(G_{0t}+G_{1}\phi_{t})+G_{0xx}+2G_{1x}\phi_{x}+G_{1}\phi_{xx}+2G_{2}
\phi_{x}^2+G_{0yy}\nonumber \\
+2G_{1y}\phi_{y}+G_{1}\phi_{yy}+2G_{2}\phi_{y}^2]
-2F_{0}[G_{0x}^2+G_{0y}^2+G_{1}^2(\phi_{x}^2+\phi_{y}^2)\nonumber \\
+2G_{1}(G_{0x}\phi_{x}+G_{0y}\phi_{y})]=0.\label{7_1b}
\end{eqnarray}\label{7_1}
\end{subequations} 
We consider two separate cases of the
manifold (i) $F_0G_0\neq -1$ (ii) $F_0G_0=-1$.
In the former case, from eqn. (\ref{7_1a}) and (\ref{7_1b}), the coefficient functions $F_2$
and $G_2$ can be expressed in terms of $F_0$, $G_0$, $F_1$ and $G_1$ 
leaving the later functions arbitrary. For the case $(1+F_0G_0)=0$, we assume for simplicity the
Kruskal's reduced manifold $\phi(x,y,t)=x+\psi(y,t)=0$.  Using this in (\ref{7_1}), we
find two sets of solutions.

{\it Case (1):}
\begin{subequations}
\begin{eqnarray}
F_{1}=\frac{i F_{0y}}{(1-i\psi_{y})}, \\
G_{1}=\frac{i G_{0y}}{(1-i\psi_{y})}.
\end{eqnarray}\label{8_1}
\end{subequations}

{\it Case (2):}
\begin{subequations}
\begin{eqnarray}
F_{1}=\frac{i F_{0y}}{(1-i\psi_{y})}, \\
G_{1}=\frac{-i G_{0y}}{(1+i\psi_{y})}.
\end{eqnarray}\label{8_2}
\end{subequations} \\
{\it 2. Resonances}

To find the resonances, that is the powers of the Laurent series
(\ref{6}) at which arbitrary functions enter,  for branches (i),
(ii) and (iii) we expand
\begin{subequations}
\begin{eqnarray}
F=F_{0}\phi^p+...+\alpha\phi^{p+r}, \\
G=G_{0}\phi^q+...+\beta\phi^{q+r},
\end{eqnarray}
\end{subequations}
($\alpha$, $\beta$ not both zero) and substitute in the equations (\ref{5}) containing the dominant
terms alone to fix the values of $r$.  Detailed calculation leads
to the following results.
\\

Branch(i) $p=-1$, $q=-1$: $r=-1,-1,0,0$

Branch(ii) $p=-1$, $q=0$: $r=-1,0,0,1$

Branch(iii) $p=0$, $q=-1$: $r=-1,0,0,1$
\\
\newline
For the case of branch (iv), $p=0$, $q=0$, we proceed with the expansion
\begin{subequations}
\begin{eqnarray}
F=F_{0}+F_1 \phi+...+F_r\phi^r, \\
G=G_{0}+G_1 \phi+...+G_r\phi^r,
\end{eqnarray}
\end{subequations}
and substitute them into the equations (\ref{5}) and collect the coefficients of
$\phi^{r-2}$ and $\phi^{r-1}$ (after making use of eqs.(\ref{7_1}). \\ \\
{\it (a) Coefficients of $\phi^{r-2}$:}

When $(1+F_0G_0)\neq0$, we have the condition
\begin{subequations}
\begin{eqnarray}
(1+F_0G_0)(1+\psi_y^2)r(r-1)F_r=0, \\
(1+F_0G_0)(1+\psi_y^2)r(r-1)G_r=0.
\end{eqnarray}\label{8_3}
\end{subequations} 
It follows that the resonance values are $r=0,0,1,1$.  For the case 
$(1+F_0G_0)=0$, the conditions become identities. \\ \\
{\it (b) Coefficients of $\phi^{r-1}$:}

When $(1+F_0G_0)\neq0$, the resulting condition is in confirmity with the
resonance values $r=0,0,1,1$ noted above.
When $(1+F_0G_0)=0$, we have
\begin{subequations}
\begin{eqnarray}
r[(F_{0}G_{1}+F_{1}G_{0})(1+\psi_{y}^2)(r-1)-4(G_{0}F_{1}(1+\psi_{y}^2)+
G_{0}F_{0y}\psi_{y})]F_{r}=0,\\
r[(F_{0}G_{1}+F_{1}G_{0})(1+\psi_{y}^2)(r-1)-4(F_{0}G_{1}(1+\psi_{y}^2)+
F_{0}G_{0y}\psi_{y})]G_{r}=0.
\end{eqnarray}
\end{subequations}
These equations reduce to the following forms for the cases 1 and 2, respectively.

{\it Case(1):}

\begin{subequations}
\begin{eqnarray}
4i r G_0 F_{0y} =0,\\
4i r F_0 G_{0y} =0.
\end{eqnarray}
\end{subequations}
For this case, the resonance values are $0,0$.

{\it Case(2):}

In this case, we have
\begin{subequations}
\begin{eqnarray}
r[F_{0y}G_0(r-5)-F_0G_{0y}(r-1)] =0,\\
r[G_{0y}F_0(r-5)-G_0F_{0y}(r-1)] =0.
\end{eqnarray}\label{8_4}
\end{subequations}
Since $(1+F_0G_0)=0$, $F_{0y}G_0+F_0G_{0y}=0$ and consequently from Eqs. (\ref{8_4}), we
find the resonance values to be $r=0,0,3,3$. \\ \\
{\it 3. Analysis of the Laurent expansion for arbitrary functions}

In the case of the branches (i), (ii) and (iii) we have verified that the
resonance conditions are indeed satisfied in the sense that apart from
the arbitrariness of the singular manifold, required number of
arbitrary functions occur at $r=0$ and $r=1$ in the Laurent series
and also that no logarithmic singularity can occur in the 
leading order for the branch (i).
We now carry out the calculations for the analysis of the Taylor
like expansion corresponding to the branch (iv) (again in terms of the
Kruskal's reduced manifold $x+\psi(y,t)=0$) by writing
\begin{subequations}
\begin{eqnarray}
F(x,y,t)=F_{0}(y,t)+F_{1}(y,t)\phi+F_{2}(y,t)\phi^2+F_{3}(y,t)\phi^3+...., \\
G(x,y,t)=G_{0}(y,t)+G_{1}(y,t)\phi+G_{2}(y,t)\phi^2+G_{3}(y,t)\phi^3+.... .
\end{eqnarray}\label{9}
\end{subequations}
Substituting the above into Eq.(\ref{5}), and collecting the
coefficients of different powers of $\phi$ we obtain the following
results. \\
\newline
{\emph{Zeroth order in $\phi$:}}

a) For the manifold $F_0G_0\neq -1$, the Taylor like series (\ref{9}) can be easily shown not to admit any movable singular manifold, where four arbitrary functions can enter into the
series (while the manifold $\phi$ can be absorbed into $F_1$ or $G_1$).
This is in confirmity with the resonance values $r=0,0,1,1$ pointed out after
Eq.(\ref{8_3}). \\

b) For the manifold $F_0G_0=-1$, one can obtain two sets of the expression 
for $F_1$ and $G_1$ which are the same as cases 1 and 2 given by Eqs. (\ref{8_1}) and (\ref{8_2}), 
respectively.  We will consider each of the cases separately. \\ \\
{\it Case(1)-Eqs.(\ref{8_1}):} \\
{\emph{(a) First order in $\phi$:}}
With $(1+F_{0}G_{0})=0$, we have
\begin{subequations}
\begin{eqnarray}
(F_{0}G_{1}+F_{1}G_{0})[i(F_{0t}+F_{1}\psi_{t})+2F_{2}+(F_{0yy}+2F_{1y}\psi_{y}
+F_{1}\psi_{yy}
\nonumber \\
+2F_2\psi_y^2)]-4G_{0}[F_{0y}F_{1y}+2F_{0y}F_{2}\psi_{y}+F_{1}(2F_{2}(1+\psi_{y}^2)+F_{1y}\psi_{y})]]
\nonumber \\
-2G_{1}[F_{1}^2(1+\psi_{y}^2)+F_{0y}(F_{0y}+2F_{1}\psi_{y})]=0, \\
(F_{0}G_{1}+F_{1}G_{0})[-i(G_{0t}+G_{1}\psi_{t})+2G_{2}+(G_{0yy}+2G_{1y}\psi_{y}
+G_{1}\psi_{yy}
\nonumber \\
+2G_2\psi_y^2)]-4F_{0}[G_{0y}G_{1y}+2G_{0y}G_{2}\psi_{y}+G_{1}(2G_{2}(1+\psi_{y}^2)+G_{1y}\psi_{y})]]
\nonumber \\
-2F_{1}[G_{1}^2(1+\psi_{y}^2)+G_{0y}(G_{0y}+2G_{1}\psi_{y})]=0. 
\end{eqnarray}
\end{subequations}
Using the results of the previous order for $F_1$ and $G_1$, we obtain
\begin{subequations}
\begin{eqnarray}
F_{2}=\frac{-F_{0yy}(1-i\psi_{y})-i F_{0y}\psi_{yy}}{2 (1-i\psi_{y})^3}, \\
G_{2}=\frac{-G_{0yy}(1-i\psi_{y})-i G_{0y}\psi_{yy}}{2 (1-i\psi_{y})^3}.
\end{eqnarray}
\end{subequations}
{\emph{(b) Second order in $\phi$:}}

Here we obtain $F_3$ and $G_3$ as
\begin{subequations}
\begin{eqnarray}
F_{3}&=&\frac{i}{12 G_0 F_{0y}}\bigg[2G_0 [4 F_2^2+(F_{1y}^2+4F_2^2\psi_y^2+4F_{1y}F_2\psi_y)
+(2F_{0y}F_{2y} \nonumber \\
& &
+F_1F_{2y}\psi_y)]+2 G_1[4F_1F_2+2(F_{0y}F_{1y}+2F_{0y}F_2\psi_y+F_1F_{1y}\psi_y
 \nonumber \\
& & +2F_1F_2\psi_y^2)]-(F_0G_2+F_1G_1+F_2G_0)[i(F_{0t}+F_1\psi_t)+2F_2
 \nonumber \\
& & +(F_{0yy}+2F_{1y}\psi_{y}
+F_{1}\psi_{yy}+2F_2\psi_y^2)]\bigg],  \\
G_{3}&=&\frac{i}{12 G_{0y} F_{0}}\bigg[2F_0 [4 G_2^2+(G_{1y}^2+4G_2^2\psi_y^2+4G_{1y}G_2\psi_y)
+(2G_{0y}G_{2y} \nonumber \\
& &
+G_1G_{2y}\psi_y)]+2 F_1[4G_1G_2+2(G_{0y}G_{1y}+2G_{0y}G_2\psi_y+G_1G_{1y}\psi_y
 \nonumber \\
 & & +2G_1G_2\psi_y^2)]-(F_0G_2+F_1G_1+F_2G_0)[-i(G_{0t}+G_1\psi_t)+2G_2
\nonumber \\
& & +(G_{0yy}+2G_{1y}\psi_{y}
+G_{1}\psi_{yy}+2G_2\psi_y^2)]\bigg].
\end{eqnarray}
\end{subequations}
In a similar way, one can compute $(F_4, G_4)$, $(F_5, G_5)$, etc.
No indeterminate coefficients appear in the series (at least upto the
order deduced) and thus no possibility for singularity arises.  We also
note that either $F_0$ or $G_0$ and $\psi(y,t)$ are the only arbitrary 
functions in the Taylor like series (\ref{9}) in confirmity with the 
resonance values $r=0,0$. \\ \\
{\it Case(2)-Eqs.(\ref{8_2}):} \\
{\emph{(a) First order in $\phi$:}}
From Eq. (\ref{5}) we obtain
\begin{subequations}
\begin{eqnarray}
F_{2}=\frac{i}{4F_{0y}G_0}\bigg(\frac{-i(F_{0y}G_0-F_0G_{0y})}{1+\psi_y^2}\bigg[
i(F_{0t}+i\frac{F_{0y}}{1-i\psi_y}\psi_t)+F_{0yy} \nonumber \\
+\frac{2iF_{0yy}}
{1-i\psi_y}\psi_y-\frac{2F_{0y}}{(1-i\psi_y)^2}\psi_y\psi_{yy}+\frac{iF_{0y}}
{1-i\psi_y}\psi_{yy} \bigg] \nonumber \\
+4G_0\bigg[\frac{F_{0y}}{1-i\psi_y}(\frac{iF_{0yy}}{1-i\psi_y}
-\frac{F_{0y}}{{(1-i\psi_y})^2}\psi_{yy})\bigg]\bigg), \\
G_{2}=\frac{-i}{4F_{0}G_{0y}}\bigg(\frac{-i(F_{0y}G_0-F_0G_{0y})}{1+\psi_y^2}\bigg[
-i(G_{0t}-i\frac{G_{0y}}{1+i\psi_y}\psi_t)+G_{0yy} \nonumber \\
-\frac{2iG_{0yy}}
{1+i\psi_y}\psi_y-\frac{2G_{0y}}{(1+i\psi_y)^2}\psi_y\psi_{yy}-\frac{iG_{0y}}
{1-i\psi_y}\psi_{yy} \bigg] \nonumber \\
+4F_0\bigg[\frac{G_{0y}}{1+i\psi_y}(\frac{-iG_{0yy}}{1+i\psi_y}
-\frac{G_{0y}}{{(1+i\psi_y})^2}\psi_{yy})\bigg]\bigg). 
\end{eqnarray}\label{9_1}
\end{subequations}
{\emph{(b) Second order in $\phi$:}}
\begin{subequations}
\begin{eqnarray}
2G_0[4F_2^2+(F_{1y}^2+4F_2^2\psi_y^2+4F_{1y}F_2\psi_y)+(2F_{0y}F_{2y}+F_1F_{2y}\psi_y)]
\nonumber \\
+2G_1[4F_1F_2+2(F_{0y}F_{1y}+2F_{0y}F_2\psi_y+F_1F_{1y}\psi_y+2F_1F_2\psi_y^2)]
 \nonumber \\
 -(F_0G_2+F_1G_1+F_2G_0)[i(F_{0t}+F_1\psi_t)+2F_2+(F_{0yy}+2F_{1y}\psi_y
 \nonumber \\
 +F_1\psi_{yy}+2F_2\psi_y^2)]-(F_0G_1+F_1G_0)[i(F_{1t}+2F_2\psi_t) \nonumber \\
+(F_{1yy}+4F_{2y}\psi_y+2F_2\psi_{yy})]=0, \\
2F_0[4G_2^2+(G_{1y}^2+4G_2^2\psi_y^2+4G_{1y}G_2\psi_y)+(2G_{0y}G_{2y}+G_1G_{2y}\psi_y)]
\nonumber \\
+2F_1[4G_1G_2+2(G_{0y}G_{1y}+2G_{0y}G_2\psi_y+G_1G_{1y}\psi_y+2G_1G_2\psi_y^2)]
 \nonumber \\
 -(F_0G_2+F_1G_1+F_2G_0)[-i(G_{0t}+G_1\psi_t)+2G_2+(G_{0yy}+2G_{1y}\psi_y
 \nonumber \\
 +G_1\psi_{yy}+2G_2\psi_y^2)]-(F_0G_1+F_1G_0)[-i(G_{1t}+2G_2\psi_t) \nonumber \\
+(G_{1yy}+4G_{2y}\psi_y+2G_2\psi_{yy})]=0.
\end{eqnarray} \label{10}
\end{subequations}
It may be noted that in this order both $F_3$ and $G_3$ are absent
indicating that they are arbitrary functions corresponding to the resonance values $r=3,3$.  
Note that from Eqs. (\ref{9_1}) and the relation $(1+F_0G_0)=0$, two of the three functions $F_0$, 
$G_0$ and $\psi$ are arbitrary corresponding to the values $r=0,0$. However,  simplifying the
above set of equations (\ref{10}) by using the expressions obtained for
the coefficients $F_1$, $F_2$, $G_1$, $G_2$ in terms of 
$F_0$, $G_0$ and $\psi$ (vide Eqs. (\ref{8_1}), (\ref{8_2}), (\ref{9_1})), 
we find that the equations (\ref{10}) reduce to two nontrivial conditions which are
incompatible, unless the $y$-dependence is dropped (corresponding
(1+1) dimensional system (\ref{1})) or one carries out
the analysis with the radial variable $r=\sqrt{x^2+y^2}$ (vide Eq.(\ref{3})). 
As a consequence logarithmic singularity appears in the series expansion
(\ref{9}).  Consequently the (2+1) dimensional continuum isotropic Heisenberg spin system (\ref{4}) and so (\ref{2})
does not satisfy the Painlev\'e property \cite{ref16} and is expected to be nonintegrable.

One can also carry out the analysis with the general manifold $\phi(x,y,t)$ instead of the 
Kruskal's reduced manifold and one can check that the same conclusion results in here also.

To conclude, in this letter we have shown that the physically
important (2+1) dimensional isotropic Heisenberg continuum spin system
(\ref{2}) does not admit the Painlev\'e property and so it belongs to the class of  non-integrable
nonlinear evolution equations.  It will be of considerable interest
to investigate the underlying spatiotemporal structures of such a nonlinear
evolution equation in detail.
\\ \\
{\bf Acknowledgement} \\
The work of C. S. and M. L. forms part of a Department of Science and
Technology, Government of India sponsored research project.


\end{document}